# Semantic Clone Detection via Probabilistic Software Modeling

Hannes Thaller[1](✉), Lukas Linsbauer[2], and Alexander Egyed[1]⋆

[1] Johannes Kepler University Linz, Austria
{hannes.thaller, alexander.egyed}@jku.at
[2] Technical University of Braunschweig, Germany
l.linsbauer@tu-braunschweig.de

**Abstract.** Semantic clone detection is the process of finding program elements with similar or equal runtime behavior. For example, detecting the semantic equality between the recursive and iterative implementation of the factorial computation. Semantic clone detection is the de facto technical boundary of clone detectors. In recent years, this boundary has been tested using interesting new approaches. This article contributes a semantic clone detection approach that detects clones which have 0 % syntactic similarity. We present Semantic Clone Detection via Probabilistic Software Modeling (SCD-PSM) as a stable and precise solution to semantic clone detection. PSM builds a probabilistic model of a program that is capable of evaluating and generating runtime data. SCD-PSM leverages this model and its model elements for finding behaviorally equal model elements. This behavioral equality is then generalized to semantic equality of the original program elements. It uses the likelihood between model elements as a distance metric. Then, it employs the likelihood ratio significance test to decide whether this distance is significant, given a pre-specified and controllable false-positive rate. The output of SCD-PSM are pairs of program elements (i.e., methods), their distance, and a decision on whether they are clones or not. SCD-PSM yields excellent results with a Matthews Correlation Coefficient greater than 0.9. These results are obtained on classical semantic clone detection problems such as detecting recursive and iterative versions of an algorithm, but also on complex problems used in coding competitions.

## 1 Introduction

Copying and pasting source code fragments leads to code clones, which are considered an anti-pattern. Code clones increase maintenance costs [31,32], promote bad software design [29,13,17], and introduce or propagate bugs [4,28,14]. However, duplicating code fragments also allows faster adaptation to requirements, the re-use of stable and well-tested solutions [25,26], and helps to overcome language limitations [21,35], thereby lowering development costs. The impact of code clones and the contradicting evidence various studies provide are the topics of an ongoing discussion in the community. Meanwhile, it is certain that developers will continue duplicating source code to leverage its benefits, despite its drawbacks. The key is the awareness and management of clones to maximize efficiency while balancing quality.

⋆ The research reported in this paper has been supported by the Austrian Ministry for Transport, Innovation and Technology, the Federal Ministry of Science, Research and Economy, and the Province of Upper Austria in the frame of the COMET center SCCH. This research was funded in part, by the Austrian Science Fund (FWF) [P25513].

Traditionally, the clone taxonomy distinguishes between four types of clones [35,2,34]. Type 1-3 describe code clones caused by copying and pasting the source code with or without changes. Type 4 clones describe code clones that do not have any syntactic similarity but implement the same functionality (semantic equivalence). For example, the recursive and iterative implementation of an algorithm (e.g., Fibonacci computation) have no syntactic similarity while implementing the same functionality. Existing tools have limited or no capabilities to detect Type 4 clones [19]. Most current studies exclude them because of the lack of tool support [23,35,2,39,11]. Nevertheless, Type 4 clones exist, and recent research efforts have tried to deepen the understanding of them [19,49,20]. This article provides a significant contribution to semantic clone detection in the form of novel concepts and a prototype implementing them.

We present *Semantic Clone Detection via Probabilistic Software Modeling (SCD-PSM)*. SCD-PSM extends our work on Probabilistic Software Modeling (PSM) [43] via a semantic clone detection pipeline. PSM builds Probabilistic Models (PMs) from programs. It analyzes the static structure and dynamic runtime behavior and replicates the program in the form of a generative probabilistic model. These models allow developers to reason about the semantics of a program. SCD-PSM extends this work by leveraging the PMs and causal reasoning to find semantically (i.e., behaviorally) equivalent code elements. SCD-PSM allows full quantification of the behavioral distance of code elements via likelihoods. Furthermore, the likelihood evaluation via PMs allows for statistical significance tests to decide whether a pair of code elements are clones. SCD-PSM detects semantic clones with no textual similarity, such as the iterative and recursive version of an algorithm. The average performance of the approach reaches a Matthews Correlation Coefficient of 0.965 on a complex problem set indicating a robust method for semantic clone detection. This work extends our previous work [41] with a full evaluation and the theoretical foundation.

Section 2 provides the background needed to understand SCD-PSM including the basics of PSM. Section 3 clarifies what semantic clones are in the context of this work. Section 4 presents the approach in which representation, search space, and the various similarity stages are described. Section 5 evaluates the approach while Section 6 discusses the results. Limitations of the approach and possible threats are given in Section 7 and Section 8. Section 9 compares the work to the state-of-art and Section 10 concludes this article.

```
1   int fa(int n){
2      product = 1
3      for(i = 1; i <= n i++)
4         product *= i
5      return product
6   }
```
Listing 1.1: *for-loop* implementation of factorial

```
1   int fb(int n){
2      product = 1
3      i = 1
4      while(i <= n)
5         product *= i
6            i++
7      return product
8   }
```
Listing 1.2: *while-loop* implementation of factorial

```
1   int fc(int n){
2      if(n <= 1) return 1
3      return fc(n - 1) * n
4   }
```
Listing 1.3: *Recursive* implementation of factorial

```
1   int fd(int n, String guard){
2      if(n < 1 && guard == "val")
3         return -1
4      if(n < 1 && guard == "throw")
5         throw Exception()
6      return fc(n)
7   }
```
Listing 1.4: *Delegate* implementation of factorial

## 2 Background

The clone detection research community has a long history and defines many concepts, algorithms, and tools. In contrast, Probabilistic Software Modeling (PSM) is relatively new and combines software engineering and probabilistic modeling. Some terms need clarification; others require an introduction if they diverge from their traditional names.

### 2.1 Clone Detection

Clone detection is the process of finding two similar program fragments. Listings 1.1 to 1.4 are four different implementations of the factorial function ($n!$). Listing 1.1 is a *for*-loop implementation, Listing 1.2 uses a *while*-loop, and Listing 1.3 is recursively defined. Finally, Listing 1.4 delegates its implementation to `fc()` from Listing 1.3 but may also return $-1$ in case of invalid inputs (including $n = 0$).

Representation, pairing, similarity evaluation, and clone decision are the core concepts of clone detection. *Representations* describe on which artifact the detector operates, such as text, graphs (e.g., AST), or probabilistic models. *Pairing* describes the selection of two code fragments that are potentially clones (e.g., `fa()` and `fb()`). Each pair is called a *candidate clone pair* (or candidate pair). The *similarity evaluation* measures the similarity of a candidate pair, e.g., by counting the number of different characters. Finally, the *clone decision* labels the candidate pair as a clone given a criterion on the similarity, e.g., less than ten different characters.

The properties of the similarity metric split clones into two groups [35]. Type 1-3 clones capture textual similarity while Type 4 clones capture semantic similarity [2,23,24,35,34,44]. Type 1 (Exact Clones) clones are program fragments that are identical except for variations in white-space and comments. Type 2 (Parameterized Clones) clones are program fragments that are structurally or syntactically similar except for changes in identifiers, literals, types, and comments. Type 3 (Near-Miss Clones) clones are program fragments that include insertions or deletions in addition to changes in identifiers, literals, types, and layouts. Type 4 (Semantic Clones) clones are program fragments that are functionally or semantically similar (i.e., perform the same computation) without textual similarities. These types are increasingly challenging to detect, with Type 4 being the most complex one. Note that the definition of *Semantic Clones* is often relaxed, where up-to 50% syntactic similarity of the code fragments is allowed (e.g., BigCloneBench [39]). However, we consider these clones as complex Type 3 clones (additions, deletions, reordering) and *not* as semantic clones. This means that semantic clones in the context of this work are clones with no syntactic similarity except for per-chance similarities.

We will use $a \simeq b$ to denote that $a$ is a clone of $b$. Furthermore, $a \not\simeq b$ denotes that $a$ is not a clone of $b$.

### 2.2 Programs & Code Elements

PSM generalizes object-oriented terms to describe *code elements* in a program. Code elements are *types* $\boldsymbol{T}$, *properties* $Pr$, and *executables* $\boldsymbol{Ex}$ that refer to, e.g., classes, fields, and methods in Java [1], or classes, properties, and functions in Python [45]. Additional code elements are *parameters* $Pr$ and *results* $Re$ of executables that refer to parameters and return values of a method. Properties, parameters, and results are *atomic* code elements that have identifiable states at runtime. Types and executables are *compositional* elements that act as a collection of atomic elements. Types *declare* properties and executables, capturing structural relationships. Executables have behavioral relationships that are categorized into *Inputs* (I) and *Outputs* (O). *Inputs* are *received parameters* $Pa^\mathcal{I}$, *read properties* $Pr^\mathcal{I}$, and *requested invocation results* $Re^\mathcal{I}$. *Outputs* are *returned executable results* $Re^\mathcal{O}$, *written properties* $Pr^\mathcal{O}$, and *provided parameters* $Pa^\mathcal{O}$. We will denote atomic elements in lowercase, and compositional elements in bold-face lowercase, e.g., $n$ and $\boldsymbol{fa}$ in Listing 1.1. Executable results are named after their executables, e.g., $fa$ in Listing 1.1. $\boldsymbol{fc} = \{n^{Pa,\mathcal{I}}, fc^{Re,\mathcal{I}}, fc^{Re,\mathcal{O}}\}$ denotes the code elements of Listing 1.3. For the sake of readability, we will omit the superscript classifiers if it is unambiguously possible, e.g., $\boldsymbol{fa} = \{n, fa\}$. The subset of *inputs* is denoted by $\boldsymbol{fc}^\mathcal{I} = \{n^{Pa,\mathcal{I}}, fc^{Re,\mathcal{I}}\}$ and *outputs* by $\boldsymbol{fc}^\mathcal{O} = \{fc^{Re,\mathcal{O}}\}$. Finally, the set of all input and output combinations is given by $bmex^{\mathcal{IO}} = \{(i, o) \in \boldsymbol{ex}^\mathcal{I} \times \boldsymbol{ex}^\mathcal{O}\}$. For example, $\boldsymbol{fd}^{\mathcal{IO}} = \{(n, fd), (guard, fd)\}$ describes the IO pairs of `fd()`.

### 2.3 Probabilistic Software Modeling

Probabilistic Software Modeling (PSM) [40] is a data-driven modeling paradigm that transforms a program into a Probabilistic Model (PM). PSM extracts the structure and behavior of a program. Code elements and their dependency graph represent the *structure* as described in Section 2.2. All observable events at runtime represent the *behavior*. The resulting PM and its model elements are a probabilistic copy of the program.

*Model elements* in the PM are the equivalent to code elements in the program. $P(x)$ denotes the probability distribution of variable $x$, e.g., $P_{fa}(n)$ denotes the probability distribution of input parameter $n$ of the fa-method. $p(x)$ denotes the probability of a specific event of a variable, e.g., $p_{fa}(n = 2)$. This extends the notation of code elements with probabilistic quantities. However, the notation reasons about the probabilistic behavior of code elements instead of their structural properties.

Each model element is a flow-based latent variable model [7] that learns an invertible mapping between the original observations and an isotropic unit norm Gaussian $\mathcal{N}(0, \mathbf{1})$ with $f : X \mapsto Z$. An example for $x \in X$ may be $n \in \boldsymbol{fa}$ with $n^z \in \boldsymbol{fa}^z$ being its latent Gaussian representation. The Gaussian latent space enables the model elements to generate new samples and evaluate the likelihood of samples.

*Generation* (or Sampling) draws, either marginally or conditionally, observations from a model element simulating the execution of the corresponding code element. For example, drawing 100 observations from $\boldsymbol{fa} \sim P_{fa}(n, fa)$, i.e., values for $n^{\mathcal{I}}$ and $fa^{\mathcal{O}}$, simulates 100 program executions of this method. An example for *conditional generation* would be $\boldsymbol{fa}_{|n<10} \sim P_{fa}(fa \mid n < 10)$ that only draws observations where $n < 10$. The process involves sampling from the latent Gaussian variables, and inverting the Gaussian samples to the original domain via the flow $f^{-1}(\boldsymbol{z}) = \boldsymbol{x}$. *Evaluation* takes observations and evaluates their likelihood under a model element. For example, $P_{fa}(n = 4, fa = 24)$ evaluates the likelihood of input 4 and output 24 under the $fa$ model element. The process of evaluation involves mapping a given sample into the latent space and evaluating it under the Gaussians $p_{\mathcal{N}(0,\mathbf{1})}(f(\boldsymbol{x}))$. Generation and evaluation are the core of any PSM applications and of SCD-PSM. A detailed description is given in our previous work [43].

## 3 Semantic Clones

A clear understanding of what SCD-PSM defines a *semantic clone* is essential in understanding the approach and its design choices.

**Definition 1.** *A semantic clone is a pair of executables whose (partial) input, and output relationships exhibit significant (conditional) similarities.*

Definition 1 defines semantic clones over the similarity between IO relationships of executables. This holds if the IO relationships are only partially similar, i.e., not all combinations of IO pairs between executables have to be similar. For example, $fd$ in Listing 1.4 has two IO pairs ($\boldsymbol{fd}^{\mathcal{IO}} = \{(n, fd), (guard, fd)\}$) while $fa$ in Listing 1.1 has one IO pair ($\boldsymbol{fa}^{\mathcal{IO}} = \{(n, fa)\}$). According to the definition, at least one IO pair comparison needs to be similar such that both executables are declared as a semantic clone (e.g., $(n, fd) \simeq (n, fa)$).

Furthermore, the similarities between IO pairs may only be conditional, i.e., the similarity of matching IO pairs might be depending on the state of any other code element in the comparison context. For example, the IO pair $(n, fd) \simeq (n, fa)$ is only a perfect clone in case that `fd.guard != "val"`. If `fd.guard == "val"`, the IO behavior would differ in case of $n = 1$ ($fd(1) \mapsto -1$ while $fa(1) \mapsto 1$). According to the definition, at least parts of the behavior need to be similar, capturing complex multidimensional behavioral patterns in IO relationships.

The rationale behind the comparison of IO relationships is one of cause and effect. If a pair of executables exhibit similar effects given similar causes, then their computational behavior is identical. Extending this rationale by multiple inputs and outputs leads to *partial conditional similarity*.

## 4 Approach

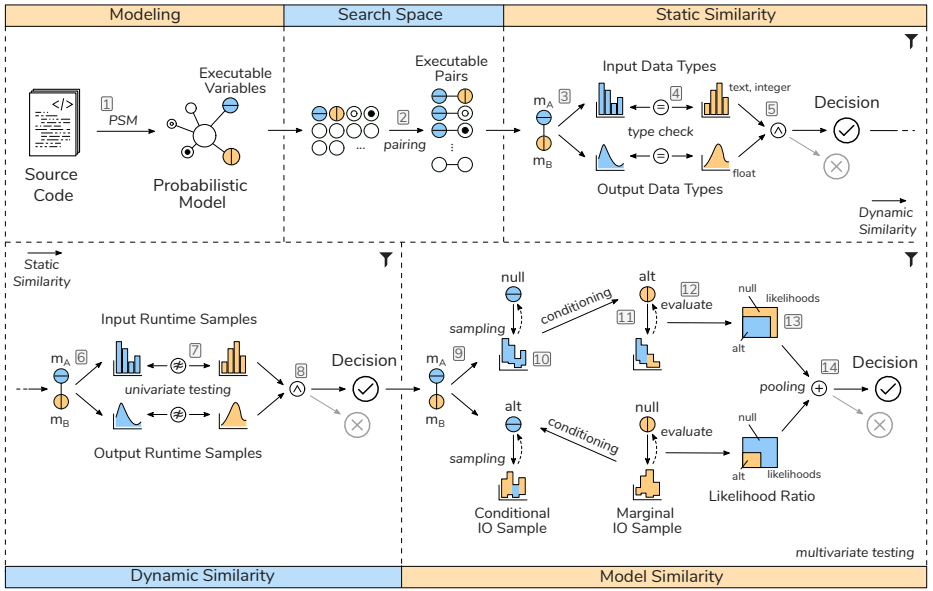

Fig. 1: The modeling phase transforms the program into a PM. The search space phase then pairs the PM model elements into candidate pairs. Finally, Static-, Dynamic- and Model Similarity evaluate the behavioral equality of the candidates.

Figure 1 illustrates SCD-PSM. It is a five-fold approach consisting of the following steps:

A. [**Modeling**] PSM builds a probabilistic model that reflects the original program;
B. [**Search Space**] A search space of candidate pairs is constructed by pairing executable model elements;

C. [**Static Similarity**] The static similarity stage accepts candidate pairs with matching data types;
D. [**Dynamic Similarity**] The dynamic similarity stage accepts candidate pairs with similar runtime data;
E. [**Model Similarity**] The model similarity stage accepts candidate pairs with similar model behavior.

The approach represents a rejecting filter pipeline that candidate pairs must traverse in order to be declared a clone. Static-, Dynamic-, and Model Similarity represent filter stages of increasing complexity.

The main contribution of this work is the implementation of a semantic clone detection pipeline on top of PSM. Further, we provide an effective process of traversing the potentially large search space of candidate pairs. Finally, we show that the behavioral equivalence of model elements generalizes to the semantic equivalence of code elements.

### 4.1 Modeling

Starting from the *Source Code* in Figure 1, PSM builds a *Probabilistic Model* (PM) [40] of the program (1). The PM is also called the Inference Graph (IG), which is a cluster graph [22] with Non-Volume Preserving Flows (NVPs) [7] as clusters. SCD-PSM uses this PM as a representation for the clone detection, similar to text-based clone detectors that use text fragments. The PM is the output of PSM and is considered as given in the context of SCD-PSM.

Executable model elements in the PM act as a surrogate to the executables in the program. SCD-PSM pairs these model elements and computes their similarity. If a behaviorally equivalent model element pair is found, then it can be seen as a semantically equivalent code element pair. In conclusion, the SCD-PSM allows for method-level semantic clone detection based on PMs representing the original executables in the program.

### 4.2 Search Space

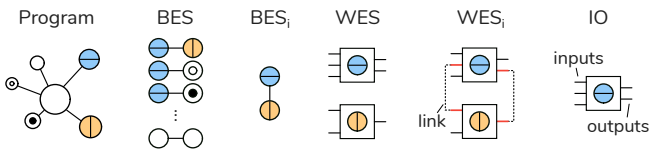

Fig. 2: SCD-PSM operates on four levels of abstraction: program, between executable, within executable, and the IO level.

SCD-PSM conducts method-level semantic clone detection, which operates on multiple abstraction levels. Figure 2 illustrates these levels, starting with the program and ending with the inputs and outputs of an executable.

The second step in Figure 1 builds a *within- and between-executable space* that SCD-PSM searches for clones. The *Between-Executable Space (BES)* is the set of executable combinations

$$BES = \{\{a, b\} \in Ex \times Ex \mid a \neq b\}, \quad (1)$$

where $exa, exb$ is a *candidate pair* (or executable pair), and $Ex$ is the set of all executables in the current analysis (illustrated in Figure 2). The theoretical size of the between-executable space are all 2-length combinations without replacement, given by

$$|BES| = \frac{|Ex|!}{2 \cdot (|Ex| - 2)!}, \quad (2)$$

where $|\cdot|$ describes the size of the underlying set. Note that the size of the BES is smaller than the Cartesian product since $\{a, b\} = \{b, a\}$. Figure 1 shows this pairing process in the Search Space aspect (2) from Figure 1. The *Within-Executable Space (WES)* is the product of IO pairs

$$WES^{ab} = \{(i, j) \in a^{\mathcal{IO}} \times b^{\mathcal{IO}}\}. \quad (3)$$

Figure 2 illustrates the WES and one IO pair from the WES that we also call *link*. The theoretical size of the within-executable space is

$$|WES^{ab}| = |a^{\mathcal{IO}}| \cdot |b^{\mathcal{IO}}| \quad (4)$$

For the sake of visualization, IO pairs are not shown in Figure 1 but are abstracted in their executable elements. The maximum theoretical search space is

$$S = \sum_i |wes(BES_i)|, \quad (5)$$

given that *wes* describes a construction function according to Equation (3), and $BES_i$ is the i'th candidate pair.

In practice, SCD-PSM evaluates only a fraction of possible combinations because of the skip evaluation. The *skip evaluation* consists of two search space limiting factors: greedy evaluation and transitive similarity. *Greedy evaluation* stops the search through the WES once a similar pair is found. The initial detection process only confirms the similarity of a candidate pair. A post-analysis can then extract all possible IO similarities for potential actions. *Transitive similarity* skips evaluations in the BES, because of $a \simeq b \simeq c$ then also $a \simeq c$ holds. In conclusion, SCD-PSM compares IO pairs of executable model elements and uses skip evaluation to traverse the search space efficiently.

### 4.3   Static Similarity

The static similarity stage is a filter that accepts candidate pairs based on their data type, as shown in Figure 1. Data types in a PSM model are integers, floats, and text.

*Input* (3) of the stage are the IO pairs $\boldsymbol{WES}^{ab} = wes(\{\boldsymbol{a}, \boldsymbol{b}\})$ of a candidate. The filter *criteria* (4) accepts a candidate pair if at least *one* link (i.e., IO pair) has a matching data type, i.e., the input but also the output have a matching data type. *Output* (5) is a boolean decision whether the candidate pair is a clone or not from a static viewpoint. If positive, then the candidate pair is moved to the next pipeline stage, i.e., the *Dynamic Similarity* evaluation (see Figure 1). If negative, then the candidate pair is marked as being *not* a clone $\boldsymbol{a} \not\simeq \boldsymbol{b}$ and no further processing is conducted. For example, the IO pairs $(n, fa) \simeq (n, fb)$ would be statically accepted as clones as both inputs and outputs have the same data type (integer). A counterexample is given by $(n, fa) \simeq (guard, fd)$ where the input data types are integers and text.

The static similarity indicates that the analyzed program is given in a programming language that allows for static analysis. Programs written in programming languages without static typing can not make use of this filter stage. In conclusion, the static similarity stage filters candidates based on their data type.

### 4.4 Dynamic Similarity

The dynamic similarity stage is a filter that accepts candidate pairs based on the runtime data, as shown in Figure 1. Candidates pairs are accepted if at least *one* IO pair (6) has an *insignificant* diverging runtime distribution (7). This boolean decision is evaluated via a Kolmogorov-Smirnov test [30], and determines whether a pair is a clone from a dynamic viewpoint (8). For example, the IO pair $(n, fa) \simeq (n, fd)$ with `guard == true` would be excluded form the filter given that runtime events with $n = 0$ reach a majority. In comparison, $(n, fa) \simeq (n, fb)$ would be accepted by the stage.

A requirement is that the candidates use a synthetic trigger. Otherwise, the comparison of the data distributions may fail because of the different modus operandi of the program. For example, running `fa` and `fb` where $n_{fa} = \mathcal{U}(0, 4)$ and $n_{fb} = \mathcal{U}(5, 10)$ would cause the dynamic stage to fail even if the implementations are equivalent. Property-based [12] or random testing can be used to generate diverse synthetic inputs.

In conclusion, the dynamic similarity stage pre-filters candidates based on univariate tests on the input and output events.

### 4.5 Model Similarity

The model similarity stage is a filter that accepts candidate pairs based on the models, as shown in Figure 1. This stage conducts a multivariate test by sampling from the executable models and cross evaluating them. This test includes the evaluation of conditional influences caused by elements that are not actively participating in an IO pair. For example, $(n, fd) \simeq (n, fa)$ holds but is conditionally dependent on *guard*. The model similarity can factor *guard* into its decision while the dynamic stage can only evaluate the average behavior of an IO pair.

*Input* (9) are the IO pairs of a candidate $\boldsymbol{WES}^{ab} = wes(\{\boldsymbol{a}, \boldsymbol{b}\})$. The crosswise log-likelihood ratio of the models is computed by *(conditional) generation* and *evaluation*. *Output* is a boolean decision on whether the candidate pair is a clone or not, from a model viewpoint. Figure 1 illustrates the entire process of the model similarity.

(A) A reference $M^{null} = \boldsymbol{a}$ and an alternative model $M^{alt} = \boldsymbol{b}$ is selected.
(B) An IO-pair $p = \boldsymbol{WES}_i^{ab}$ is selected as the target of the comparison (link).
(C) A reference sample $D^{null}$ is drawn from $M^{null}$ (10).
(D) An alternative sample $D^{alt|null}$ is drawn from $M^{alt}$ by optimizing towards the $p$ dimensions in the $D^{null}$, effectively conditioning the drawn samples (11).
(E) $D^{null}$ is evaluated under $M^{null}$ resulting the reference log-likelihood $LL^{null}$
(F) $D^{alt|null}$ is evaluated under $M^{alt}$ (12) yielding the alternative log-likelihood $LL^{null}$.
(G) Finally, the likelihood ratio of the link is computed by $\lambda = LL^{alt} - LL^{null}$

The roles between the *null* and *alt* models are then swapped, and the process is repeated. Both log-likelihood ratios are then combined by a pooling operator to produce the clone decision (14).

The role-swap is needed to avoid sub-model relationships. For example, if $M^{null} = \mathcal{N}(0,3)$ and $M^{alt} = \mathcal{N}(0,1)$ then $LL^{alt}$ will be very high because $M^{alt}$ is a sub-model from $M^{null}$. Reversing the roles highlights the differences in the models.

The final decision is based on the Generalized Likelihood Ratio Test (GLRT) [10]. It measures whether the log-likelihoods are significantly different from 0, where $\lambda$ is the test statistic. The null hypothesis is rejected for small ratios $\lambda \leq c$ where $c$ is set to an appropriate false-positive rate. For example, $\lambda < \log(0.01)$ allows 1 out of 100 candidates to be a false-positive, i.e., wrongly rejecting semantic equivalence. The pooling operator combines the link results either via hard or soft pooling. *Hard pooling* conducts for both links a GLRT yielding a positive decision if *both* links are positive. *Soft pooling* averages the link log-likelihoods ratios and then computes the GLRT yielding a positive decision if the joint GLRT is positive. Hard pooling does not allow any sub-model relationships, while soft pooling relaxes this constraint.

In conclusion, the model similarity conducts a multivariate significance test between two models, including possible conditional dependencies.

## 5  Study

This study answers the following research questions.

Q1 Does behavioral equality between model elements generalize to semantic equality of code elements?
Q2 Does the skip evaluation significantly reduce the computational demand of SCD-PSM?
Q3 Does the skip evaluation negatively impact the detection performance (i.e., precision, recall, and MCC)?

Q1 answers whether semantic clones can be detected via SCD-PSM. Q2 answers whether the search space can be efficiently processed using skip evaluation. Q3 answers how the skip evaluation influences the performance of the detection process. This is important because candidate pairs might be skipped based on false-positives or false-negatives.

### 5.1 Setup

We implemented a prototype for SCD-PSM on top of Gradient [40], a prototype for PSM. The elements and data flow of the detection process are shown in Figures 1 and 2.

1. The input *Source Code* were 13 different clone classes with a total of 108 implementation variants. This includes classical algorithms implemented recursively and iteratively such as bubble sort, as well as hard problems from the programming competition Google Code Jam[1].
2. The *Probabilistic Model* was computed via Gradient, a PSM prototype. We used the same hyper-parameters as reported in our previous work [43].
3. The *Search Space*, i.e., the *BES* and *WES*, was created according to Section 4.2 based on *all* available examples.
4. Each valid candidate pair was then submitted to the *Static-, Dynamic*, - and *Model-Similarity* stages and filtered according to Sections 4.3 to 4.5. Candidates that passed the entire filter pipeline were marked as clones.

### 5.2 Dataset

The study uses three well-known algorithms and 10 Google Code Jam 2017 (GCJ)[1] problems. The total dataset contains 108 implementation variants across 13 clone classes described by *Instance*.

Each clone class was differentially tested to verify the behavior across instances. Factorial, Fibonacci, and Sort do not need any further explanation. The GCJ problems are well specified complex optimization problems packaged in an everyday theme.

The dataset contains in total 5778 (see Equation (2)) candidate pairs of which 458 are semantic clones and 5320 are not. This yields a positive to negative ratio of 1 : 11.6, indicating a highly imbalanced distribution. An even more pronounced imbalance is to be expected in real-world applications.

Each instance was triggered with input data to allow PSM to model the different implementations. Factorial, Fibonacci, and Sort were triggered by sampling from a uniform distribution $\mathcal{U}(0, 20)$. GCJ problems were triggered by the input data provided by the competition. Each instance received the same trigger.

GCJ problems read from and write to the standard stream, which is impractical in terms of reproducibility. Our dataset is constructed such that each implementation has a `run`-method representing the cloned executable. The study results are limited to the `run`-method even if the solutions use helper methods.

---

[1] https://codingcompetitions.withgoogle.com/codejam/archive

Helper methods may, for example, be methods that compute parts of the final solution, or reorganize the data. This guarantees a proper problem scope, a well-defined recall and precision, and a clearly defined benchmark for future reproducibility.

### 5.3 Controlled Variables

The study controls for the search space *Evaluation* strategy, *Dynamic False-Positive Rate (D-FPR)*, *Model False-Positive Rate (M-FPR)*, and *Pooling*.

**Evaluation** describes how the search space is processed: *exhaustive*, or *skip*. The exhaustive evaluation compares each executable candidate with each other. The skip evaluation uses the transitive similarity (see Section 4.2) and may skip evaluations if possible.

**Dynamic False-Positive Rate (D-FPR)** defines the critical value $\alpha$ of the Kolmogorov-Smirnov test with 0.001 and 0.01, at which similarity is rejected.

**Model False-Positive Rate (M-FPR)** defines the critical value $c$ of the Generalized Likelihood Ratio test with 0.001 and 0.01, at which similarity is rejected.

**Pooling** defines how the likelihood ratios from the two link directions are combined (see Figure 1, (8)) with values: *hard*, or *soft*. *Hard* pooling evaluates whether each link reaches the critical value $c$ and accepts the clone if both links evaluate as positive with $\lambda_{Link_A} \leq \frac{\log c}{2}$ and $\lambda_{Link_B} \leq \frac{\log c}{2}$. *Soft* pooling evaluates the average log-likelihood ratios (geometric mean of likelihoods) $\frac{\lambda_{Link_A} + \lambda_{Link_B}}{2} \leq \log c$, and compares it against the critical value $c$.

An additional fixed parameter is the *number of particles*. It defines the sample size that is generated during the model similarity $|D| = 50$.

### 5.4 Response Variables

The response measures of the study are the number of *Skip Evaluations*, processing *Duration*, *TP*, *FP*, *TN*, *FN*, *Precision*, *Recall*, *F1*, and *Matthews Correlation Coefficient*.

**Skip Evaluations** measures the number of evaluations that were skipped due to the skip evaluation strategy.

**Duration** measures the elapsed time to compute one candidate pair.

**TP, FP, TN, FN** measures the True Positive (TP), False Positive (FP), True Negative (TN), and False Negative (FN) detection results compared to the ground truth.

**Precision** measures the fraction of detected clones that are truly clones.

**Recall** measures the fraction of semantic clones that have been found.

**F1** measures the accuracy of a binary classification as the harmonic mean of recall and precision.

Table 1: Results of the top-5 and bottom-1 experiment along with the average performance of the top-5.

| | Controlled Variables | | | | Response Variables | | | | | | | | |
|---|---|---|---|---|---|---|---|---|---|---|---|---|---|
| Nr | Evaluation | D-FPR | M-FPR | Pooling | Duration | TP | FP | TN | FN | Skip | Precision | Recall | F1 | **MCC** |
| 1 | skip | 0.100 | 0.001 | soft | 1560 | 437 | 0 | 5320 | 21 | 345 | 1.000 | 0.954 | 0.977 | **0.975** |
| 2 | skip | 0.010 | 0.001 | soft | 1620 | 437 | 0 | 5320 | 21 | 345 | 1.000 | 0.954 | 0.977 | **0.975** |
| 3 | exhaustive | 0.010 | 0.001 | soft | 1680 | 425 | 0 | 5320 | 33 | 0 | 1.000 | 0.928 | 0.963 | **0.960** |
| 4 | skip | 0.010 | 0.010 | soft | 1920 | 423 | 0 | 5320 | 35 | 332 | 1.000 | 0.924 | 0.960 | **0.958** |
| 5 | exhaustive | 0.100 | 0.001 | soft | 2040 | 421 | 0 | 5320 | 37 | 0 | 1.000 | 0.919 | 0.958 | **0.955** |
| 16 | exhaustive | 0.100 | 0.010 | hard | 2820 | 293 | 0 | 5320 | 165 | 0 | 1.000 | 0.639 | 0.780 | **0.787** |
| 1-5 | skip | 0.010 | 0.001 | soft | 1740 | 428 | 0 | 5320 | 29 | 340 | 1.000 | 0.936 | 0.967 | **0.965** |

Duration in seconds

**Matthews Correlation Coefficient (MCC)** measures the quality of the clone detection in the form of a correlation ranging from $-1$ to $1$, with $0$ being a random selection. The MCC will be the reference performance metric as it is the most robust metric in an imbalanced binary classification setting [3]. It is a correlation coefficient which may be interpreted by the guidelines proposed by Evans [9].

## 5.5 Comparison of Clone Detectors

In total, eight alternative approaches are used to contextualize the performance of SCD-PSM. The alternatives have a wide variety in terms of internal representation and clone detection capabilities as listed in Table 3. ASTNN (8) and ASTNN Leaky (9) are the same approach but have different evaluation methods. ASTNN Leaky (9) uses a random split of the dataset as reported by the authors [50]. It overestimates the performance of the approach via a lack of isolation between training and test dataset. For example, $fa \simeq fb$ and $fa \simeq fc$ might be in the train split while $fb \simeq fc$ might be in the test split. ASTNN (8) uses a group-wise Cross Validation (CV), where clone classes are entirely isolated either into the training or test proportion of the dataset. This represents a real-world situation where first the detector is fitted and then applied to a new system with unknown code fragments.

Detectors that report lines instead of methods may produce more results (TP, FP, TN, FN) than present in the dataset. A similar situation is given by ASTNN Leaky that runs multiple evaluations via the cross validation.

## 5.6 Experiment Results

Creating the PSM model with Gradient took 2134.38 s, resulting in an average modeling time of 19.75 s for the 195 executables. This includes 87 helper methods.

Table 1 contains the aggregate results of the top-5 experiments along with the results of the worst experiment. The bottom line in Table 1 is the average

Table 2: Performance breakdown of the best performing experiment listed as Nr. 1 in Table 1.

| Stage | Duration | TP | FP | TN | FN | Precision | Recall | F1 | **MCC** |
|---|---|---|---|---|---|---|---|---|---|
| initial | –      | 458 | 5320 | 0    | 0  | 0.079 | 1.000 | 0.147 |       |
| static  | 0.0001 | 458 | 1504 | 3816 | 0  | 0.233 | 1.000 | 0.379 | **0.409** |
| dynamic | 0.208  | 451 | 50   | 5270 | 7  | 0.900 | 0.985 | 0.941 | **0.936** |
| model   | 1.749  | 437 | 0    | 5320 | 21 | 1.000 | 0.954 | 0.977 | **0.975** |
|         | 0.344  | 437 | 0    | 5320 | 21 | 0.996 | 0.954 | 0.977 | **0.975** |

Duration in seconds

performance of the top-5 experiments. The generally expected performance of the approach is *very strong* with an MCC of 0.965. High confidence for negative examples is given with no false-positives reflecting the pipeline's FPR rates (D-FPR × M-FPR). The best experiment featured a *skip evaluation*, *0.100* D-FPR and *0.001* M-FPR rates, and *soft pooling* (Nr. 1) with an MCC of 0.975. The worst experiment featured a *exhaustive evaluation*, D-FPR of 0.100, M-FPR of 0.010, and *hard pooling* (Nr.16) with a *strong* MCC of 0.787. A total of 345 candidates were skipped while reaching a recall of 0.933.

Table 2 lists the cumulative performance of the best model, starting with an initial prediction that all candidates are semantic clones (rejecting pipeline). The *static* stage finds 71.729 % (3816) of the FPs, improving the MCC by 0.409. The *dynamic* stage additionally removes another 27.330 % (1454) of FPs but introduces 1.528 % (7) of the possible FNs. An improvement of the MCC by 0.527 is achieved via the dynamic stage. Finally, the *model* stage removes the remaining 0.939 % (50) FPs but introduces additional 3.056 % (14) additional FNs. The model stage improves the MCC by 0.039.

On average, 5.884 % (340) of the total 5778 evaluations could be skipped. This equals 74.235 % of the total 458 TPs. On average 37.359 % (50 354) of the total 134 782 IO pair evaluations could be saved via greedy evaluation. The average duration of the exhaustive experiments was 2394 s, leading to 414 ms per candidate. Skip experiments lasted on average 1988 s with 344 ms per candidate. The static stage lasted on average for <0.001 % of the time per candidate (see Table 2), the dynamic stage for 0.106 %, and the model stage for 0.893 %.

Table 3 lists the detection results of eight alternative clone detectors. Simian, NiCad, and CCAligner found no clones in the dataset. PMD, SourcererCC, Oreo, and iClones found some clones (< 20) with a low recall (4 %). Each of these detectors has a *very weak* performance below an MCC of 0.20  ASTNN with the leaky evaluation has a *very strong* performance with an MCC of 0.976. ASTNN 3-Group CV has a *strong* performance with an MCC of 0.711. The longest computational duration is given by ASTNN with 1034 min.

Table 3: Detection results of other clone detectors on the dataset.

| Nr | Tool | Note | Repr. | Type | Duration | TP | FP | TN | FN | Precision | Recall | F1 | MCC |
|---|---|---|---|---|---|---|---|---|---|---|---|---|---|
| 1 | Simian [16] | | Text | 1 | 0.138 | 0 | 0 | 5320 | 458 | | 0.000 | | |
| 2 | NiCad [5] | | Text | 3 | 1.291 | 0 | 0 | 5320 | 458 | | 0.000 | | |
| 3 | CCAligner [47] | | Token | 3 | 1.109 | 0 | 4 | 5316 | 458 | 0.000 | 0.000 | | **-0.007** |
| 4 | PMD [33] | | Token | 2 | 1.389 | 8 | 12 | 5308 | 450 | 0.400 | 0.017 | 0.033 | **0.069** |
| 5 | SourcererCC [37] | | Token | 3/4 | 36.86 | 10 | 0 | 5320 | 448 | 1.000 | 0.021 | 0.042 | **0.142** |
| 6 | Oreo [36] | | Model | 3/4 | 79.00 | 17 | 5 | 5315 | 441 | 0.772 | 0.037 | 0.070 | **0.158** |
| 7 | iClones [15] | | Token | 3/4 | 0.980 | 13 | 0 | 5320 | 445 | 1.000 | 0.028 | 0.055 | **0.161** |
| 8 | ASTNN [50] | 3-Group CV | Model | 4 | 1034 | 296 | 29 | 1415 | 162 | 0.911 | 0.646 | 0.756 | **0.711** |
| 9 | ASTNN (Leaky) | Random Split | Model | 4 | 2028 | 442 | 4 | 5316 | 16 | 0.991 | 0.965 | 0.978 | **0.976** |
| 10 | SCD-PSM | Top 1-5 | Model | 4 | 1740 | 428 | 0 | 5320 | 29 | 1.000 | 0.936 | 0.967 | **0.965** |

Duration in seconds

## 6 Discussion

The goal of the study was to provide evidence of whether behavioral equality of model elements generalizes to semantic equality of code elements (Q1). Furthermore, we were interested in the skip evaluation and its performance implications (Q2 and Q3).

### 6.1 Research Question 1 — Detection Performance

Table 1 and Table 2 present strong results in favor of Q1. The MCC for the top-5 experiments was *very strong* with all MCCs being above 0.9. Even the worst experiment still yielded a *moderate* performance of 0.749.

Table 3 provides additional context to the results by presenting the detection results of alternative clone detectors. As expected, tools relying heavily on the textual representation of clones have very low recall (Simian, NiCad, CCAligner, PMD) on the dataset. Most clones found by the alternative tools span only a few lines of code. In contrast, iClones finds large clones that include array accesses and manipulations. ASTNN is the best comparison tool and finds many clones with good precision. The approach is sensitive to hyper-parameters and to the training and test split, leading in some cases to a test performance close to MCC of 0. The low recall for Type 1-3 detectors indicate the high quality of the dataset. The moderate recall for Type 3/4 detectors indicate the high quality of SCD-PSM. Given this evidence, we conclude that Q1 holds.

> Q1 — Behavioral equality between model elements generalizes to semantic equality of code elements, allowing for semantic clone detection via probabilistic software modeling.

## 6.2  Research Question 2 — Skip Evaluation Scalability

The goal of the static and dynamic stage is to reduce the number of evaluations that the model stage must conduct. Each stage incurs an increasing cost of evaluation per candidate, with the model stage taking the largest share of the evaluation time, 89 %. Every TP has to pass the model stage to be declared a clone (rejecting pipeline). The skip evaluation avoided, on average, the re-computation of 74 % (340) of the TP candidate pairs. The greedy evaluation avoided, on average, the evaluation of 37 % of IO pairs. This offloads most of the evaluation time to the earlier stages, which are computationally inexpensive, while shortcutting the model stage. In comparison to the alternative detectors, SCD-PSM needs substantially more time to compute (1.32 min vs. 29 min). An exception is ASTNN which has a similar runtime as SCD-PSM. Most of the runtime of SCD-PSM is caused by the operational overhead, e.g., loading the model from the database. Optimizing this overhead, as a theoretical maximum, could reduce the overall runtime on the dataset to 6.49 min given the average durations for each stage in Table 2. In conclusion, the skip evaluation reduces the number of model evaluations, which are responsible for most of the evaluation time, down to a quarter.

> Q2 — Skip evaluation reduces the number of evaluations for the most expensive stage (model) in the SCD-PSM pipeline significantly.

## 6.3  Research Question 3 — Skip Evaluation Effects

Skip evaluation can cause cascading errors given an FP. Once an FP is introduced, every semantic clone related to the FP has a chance to become an FP in the same (wrong) clone class itself. These cascading FPs are potential sources of serious performance degradation. Skip evaluation experiments are ranked higher and are significantly better than experiments that conducted an exhaustive search. However, the absolute performance gain is only a MCC of 0.056, hinting at a per-chance significance introduced by the small sample size (16 experiments). Nevertheless, given the evidence in Table 1 and Section 5.6, we can conclude that skip evaluation does not affect the performance of the detector.

> Q3 — The skip evaluation has no negative impact on the performance of the detector given low false-positive rates.

## 7  Limitations

SCD-PSM inherits the limitations of PSM, such as its need for a runnable program to build the model. PSM only models the application structure and its data, not references. References are changing addresses with no relation to the running program. Hence, they have no meaningful underlying distribution that can be modeled. However, once references are dereferenced, e.g., by accessing a field, their accessed data will be part of the model and therefore usable in SCD-PSM. Nevertheless, algorithms with the sole purpose of manipulating references do not work with SCD-PSM.

PSM explodes lists into singular values, since distributions do not contain any order information. This means executables that change the order of sequences are matched based on the values, not their order. As a consequence, an ascending and descending sorting algorithm are semantically equivalent, leading to a false-positive. Extending PSM to distributions of sequences alleviates the issue but is not a trivial task.

SCD-PSM cannot detect Type 2-3 clones since textual similarities represent a different problem set. A proof can easily be constructed by adding an arbitrary number of statements that do not influence the behavior of the program but mislead text based detectors. Inversely, changing one character, e.g., a multiplication to a division, may alter the entire behavior while preserving the general textual similarity.

We employed a controlled laboratory evaluation strategy that allowed us to exactly evaluate the performance metrics and fairly compare them between different clone detectors. This follows a recent trend [38,46,48] in the light of some criticism of opportunistic evaluations on arbitrary open source projects. The controlled laboratory evaluation provides purely functional performance results given a fixed and controlled sample of programs. The generalizability of results obtained from laboratory evaluations is limited; Using an opportunistic evaluation strategy avoids this problem. However, the strategy is prone to biases caused by the human oracles (often the authors themselves) or proxy oracles that evaluate the clones. Moreover, a fair comparison between detectors is hardly possible because the true recall of clones is in general unknown. A combination of both evaluation strategies may yield precise and generalizable results. The extension to this study is part of our future work.

## 8  Threats to Validity

A threat to validity in any semantic clone detection study is given by the programs and code fragments used in the evaluation. Semantic clones may not exhibit the same functional behavior or share too many lexicographical similarities. This study tested every clone class on its behavioral equality. Furthermore, we evaluated text-, token-, graph- and model-based detectors capable of detecting Type 1-3 clones. The low performance of Type 1-3 detectors confirmed the high quality of semantic clones in the benchmark.

## 9  Related Work

We started this article by defining what *semantic clones* means in the context of our approach (Section 3). While our definition is motivated by the capabilities of our approach, we can see strong similarities to the definition of Juergens [19]. Both definitions define behavioral similarity via IO relationships. Also, Juergens already discussed a notion of partial and conditional similarity. This understanding of Type 4 clones can be seen in multiple more recent studies [8,6,27]. In that, we see the progress of the community in terms of Type 4 clones as the definition becomes more specific.

Many studies evaluated textual clones. However, only a few studies have reported results on semantic clones without relaxing the definition of Type 4. Rattan [34] et al. provided a review of clone detection studies including approaches focused on Type 4 clones. They concluded that some approaches solve approximations (i.e., complex Type 3 clones) of Type 4 clones.

Test-based methods randomly trigger the execution of candidates and measure whether equal inputs cause similar outputs. Jiang and Su [18] were able to find semantically equivalent methods without any syntactic similarities. A similar approach was presented by Deissenboeck et al. [6]. One issue with test-based clone detection is that candidates need a similar signature. Differences in data types or the number of parameters can not be effectively handled. SCD-PSM works similarly to test-based methods in that it observes the runtime and compares the resulting behavior. However, SCD-PSM builds generative models from the observed behavior, capable of generating, conditioning, and evaluating data. This allows SCD-PSM to bridge signature mismatches by imputing missing code elements and the using a generalized type system.

Zhao and Huang [51] developed DeepSim, which phrases the problem as a binary classification task. DeepSim uses neural networks to learn encodings of the control and data flow without observing the program's runtime. PSM also uses neural networks but learns an underlying representation of the data flow and runtime. DeepSim was also evaluated on a Google Code Jam dataset. It reached an F1 score of 0.76 on the GCJ 2016 competition, while SCD-PSM reached 0.967 on the GCJ 2017. While not entirely comparable, the results are a good approximation given the similarity in the datasets.

## 10   Conclusions and Future Work

In this article, we presented Semantic Clone Detection via Probabilistic Software Modeling (SCD-PSM). PSM builds a Probabilistic Model (PM) from a program that can be used to simulate or evaluate a program. We used these PMs to detect semantic clones in programs that have 0 % syntactic similarity.

We discussed the representation, search space, static-, dynamic-, and model-similarity stages forming the main aspects of SCD-PSM. The study evaluated SCD-PSM in great detail resulting in an average MCC greater than 0.9. Also, the study showed the capability to control the false-positive rate, which is important for an industry adoption. Finally, we concluded that behavioral equality of model elements generalizes to semantic equality of code elements.

Our future work focuses on constructing a comprehensive benchmark covering controlled and real-world systems for improved generalizability of clone detection studies. Furthermore, semantic clone detection has the potential to enable new methods for fault localization applications [42].

# References

1. Arnold, K., Gosling, J., Holmes, D.: The Java Programming Language. Addison-Wesley Longman Publishing Co., Inc., Boston, MA, USA, 3rd edn. (2000)
2. Bellon, S., Koschke, R., Antoniol, G., Krinke, J., Merlo, E.: Comparison and Evaluation of Clone Detection Tools. IEEE Transactions on Software Engineering **33**(9), 577–591 (2007). https://doi.org/10.1109/TSE.2007.70725
3. Boughorbel, S., Jarray, F., El-Anbari, M.: Optimal classifier for imbalanced data using matthews correlation coefficient metric. PloS one **12**(6), e0177678 (2017)
4. Chou, A., Yang, J., Chelf, B., Hallem, S., Engler, D.: An empirical study of operating systems errors. ACM SIGOPS Operating Systems Review **35**(5), 73 (Dec 2001). https://doi.org/10.1145/502059.502042
5. Cordy, J.R., Roy, C.K.: The NiCad Clone Detector. In: 2011 IEEE 19th International Conference on Program Comprehension. p. 219–220 (Jun 2011). https://doi.org/10.1109/ICPC.2011.26
6. Deissenboeck, F., Heinemann, L., Hummel, B., Wagner, S.: Challenges of the Dynamic Detection of Functionally Similar Code Fragments. In: 2012 16th European Conference on Software Maintenance and Reengineering. p. 299–308 (Mar 2012). https://doi.org/10.1109/CSMR.2012.38
7. Dinh, L., Sohl-Dickstein, J., Bengio, S.: Density estimation using Real NVP. arXiv:1605.08803 [cs, stat] (May 2016)
8. Elva, R., Leavens, G.T.: JSCTracker : A Semantic Clone Detection Tool for Java Code (2012)
9. Evans, J.D.: Straightforward Statistics for the Behavioral Sciences. Brooks/Cole Pub. Co, Pacific Grove (1996)
10. Fan, J., Zhang, C., Zhang, J.: Generalized Likelihood Ratio Statistics and Wilks Phenomenon. The Annals of Statistics **29**(1), 153–193 (2001)
11. Farmahinifarahani, F., Saini, V., Yang, D., Sajnani, H., Lopes, C.V.: On Precision of Code Clone Detection Tools. In: 2019 IEEE 26th International Conference on Software Analysis, Evolution and Reengineering (SANER). p. 84–94 (Feb 2019). https://doi.org/10.1109/SANER.2019.8668015
12. Fink, G., Bishop, M.: Property-based testing: A new approach to testing for assurance. ACM SIGSOFT Software Engineering Notes **22**(4), 74–80 (Jul 1997). https://doi.org/10.1145/263244.263267
13. Fowler, M., Beck, K.: Refactoring: Improving the Design of Existing Code. The Addison-Wesley Object Technology Series, Addison-Wesley, Reading, MA (1999)
14. Geiger, R., Fluri, B., Gall, H.C., Pinzger, M.: Relation of Code Clones and Change Couplings. In: Baresi, L., Heckel, R. (eds.) Fundamental Approaches to Software Engineering, vol. 3922, p. 411–425. Springer Berlin Heidelberg, Berlin, Heidelberg (2006). https://doi.org/10.1007/11693017_31
15. Göde, N., Koschke, R.: Incremental Clone Detection. In: 2009 13th European Conference on Software Maintenance and Reengineering. p. 219–228 (Mar 2009). https://doi.org/10.1109/CSMR.2009.20
16. Harris, S.: Simian - Similarity Analyser (2003)
17. Hunt, A., Thomas, D.: The Pragmatic Programmer: From Journeyman to Master. Addison-Wesley, Reading, Mass (2000)
18. Jiang, L., Su, Z.: Automatic Mining of Functionally Equivalent Code Fragments via Random Testing. In: Proceedings of the Eighteenth International Symposium on Software Testing and Analysis. p. 81–92. ISSTA '09, ACM, New York, NY, USA (2009). https://doi.org/10.1145/1572272.1572283

19. Juergens, E., Deissenboeck, F., Hummel, B.: Code Similarities Beyond Copy & Paste. In: 2010 14th European Conference on Software Maintenance and Reengineering. p. 78–87. IEEE, Madrid (Mar 2010). https://doi.org/10.1109/CSMR.2010.33
20. Kafer, V., Wagner, S., Koschke, R.: Are there functionally similar code clones in practice? In: 2018 IEEE 12th International Workshop on Software Clones (IWSC). p. 2–8. IEEE, Campobasso (Mar 2018). https://doi.org/10.1109/IWSC.2018.8327312
21. Kapser, C.J., Godfrey, M.W.: "Cloning considered harmful" considered harmful: Patterns of cloning in software. Empirical Software Engineering **13**(6), 645–692 (Dec 2008). https://doi.org/10.1007/s10664-008-9076-6
22. Koller, D., Friedman, N.: Probabilistic Graphical Models: Principles and Techniques. Adaptive Computation and Machine Learning, MIT Press, Cambridge, MA (2009)
23. Koschke, R.: Survey of research on software clones. In: Koschke, R., Merlo, E., Walenstein, A. (eds.) Duplication, Redundancy, and Similarity in Software. No. 06301 in Dagstuhl Seminar Proceedings, Internationales Begegnungs- und Forschungszentrum für Informatik (IBFI), Schloss Dagstuhl, Germany, Dagstuhl, Germany (2007)
24. Krinke, J.: Identifying Similar Code with Program Dependence Graphs. Proceedings Eighth Working Conference on Reverse Engineering p. 301–309 (2001). https://doi.org/10.1109/WCRE.2001.957835
25. Krinke, J.: Is Cloned Code More Stable than Non-Cloned Code? Proceedings - 8th IEEE International Working Conference on Source Code Analysis and Manipulation, SCAM 2008 p. 57–66 (2008). https://doi.org/10.1109/SCAM.2008.14
26. Krinke, J.: Is Cloned Code Older than Non-Cloned Code? (2011)
27. Li, G., Liu, H., Jiang, Y., Jin, J.: Test-Based Clone Detection: An Initial Try on Semantically Equivalent Methods. IEEE Access **6**, 77643–77655 (2018). https://doi.org/10.1109/ACCESS.2018.2883699
28. Li, Z., Lu, S., Myagmar, S., Zhou, Y.: CP-Miner: Finding Copy-Paste and Related Bugs in Large-Scale Software Code. IEEE Transactions on Software Engineering **32**(3), 176–192 (2006). https://doi.org/10.1109/TSE.2006.28
29. Martin, R.C. (ed.): Clean Code: A Handbook of Agile Software Craftsmanship. Prentice Hall, Upper Saddle River, NJ (2009)
30. Massey, F.J.: The Kolmogorov-Smirnov Test for Goodness of Fit. Journal of the American Statistical Association **46**(253), 68–78 (Mar 1951). https://doi.org/10.1080/01621459.1951.10500769
31. Mayrand, Leblanc, Merlo: Experiment on the automatic detection of function clones in a software system using metrics. In: Proceedings of International Conference on Software Maintenance ICSM-96. p. 244–253. IEEE, Monterey, CA, USA (1996). https://doi.org/10.1109/ICSM.1996.565012
32. Monden, A., Nakae, D., Kamiya, T., Sato, S., Matsumoto, K.: Software quality analysis by code clones in industrial legacy software. In: Proceedings Eighth IEEE Symposium on Software Metrics. p. 87–94. IEEE Comput. Soc, Ottawa, Ont., Canada (2002). https://doi.org/10.1109/METRIC.2002.1011328
33. PMD: Pmd. PMD (2019)
34. Rattan, D., Bhatia, R., Singh, M.: Software clone detection: A systematic review. Information and Software Technology **55**(7), 1165–1199 (Jul 2013). https://doi.org/10.1016/j.infsof.2013.01.008
35. Roy, C.K., Cordy, J.R.: A Survey on Software Clone Detection Research. Queen's School of Computing TR **115**, 115 (2007)
36. Saini, V., Farmahinifarahani, F., Lu, Y., Baldi, P., Lopes, C.V.: Oreo: Detection of clones in the twilight zone. In: Proceedings of the 2018 26th ACM Joint Meeting on European Software Engineering Conference and Symposium on the Foundations of Software Engineering - ESEC/FSE 2018. p. 354–365. ACM Press, Lake Buena Vista, FL, USA (2018). https://doi.org/10.1145/3236024.3236026

37. Sajnani, H., Saini, V., Svajlenko, J., Roy, C.K., Lopes, C.V.: Sourcerercc: Scaling code clone detection to big-code. In: Proceedings of the 38th International Conference on Software Engineering. p. 1157–1168 (2016)
38. Su, F.H., Bell, J., Harvey, K., Sethumadhavan, S., Kaiser, G., Jebara, T.: Code relatives: detecting similarly behaving software. In: Proceedings of the 2016 24th ACM SIGSOFT International Symposium on Foundations of Software Engineering - FSE 2016. ACM Press (2016). https://doi.org/10.1145/2950290.2950321
39. Svajlenko, J., Roy, C.K.: Evaluating clone detection tools with BigCloneBench. In: 2015 IEEE International Conference on Software Maintenance and Evolution (ICSME). p. 131–140. IEEE, Bremen, Germany (Sep 2015). https://doi.org/10.1109/ICSM.2015.7332459
40. Thaller, H., Linsbauer, L., Egyed, A.: Feature Maps: A Comprehensible Software Representation for Design Pattern Detection. In: 2019 IEEE 26th International Conference on Software Analysis, Evolution and Reengineering (SANER). p. 207–217. IEEE, Hangzhou, China (Feb 2019). https://doi.org/10.1109/SANER.2019.8667978
41. Thaller, H., Linsbauer, L., Egyed, A.: Towards Semantic Clone Detection via Probabilistic Software Modeling. In: 2020 IEEE 14th International Workshop on Software Clones (IWSC). p. 64–69. IEEE (2020)
42. Thaller, H., Linsbauer, L., Egyed, A., Fischer, S.: Towards Fault Localization via Probabilistic Software Modeling. In: 2020 IEEE 3rd International Workshop on Validation, Analysis, and Evolution of Software Tests (VST). p. 24–27. IEEE (2020)
43. Thaller, H., Linsbauer, L., Ramler, R., Egyed, A.: Probabilistic Software Modeling: A Data-driven Paradigm for Software Analysis. arXiv:1912.07936 [cs] (Dec 2019)
44. Thaller, H., Ramler, R., Pichler, J., Egyed, A.: Exploring code clones in programmable logic controller software. In: 2017 22nd IEEE International Conference on Emerging Technologies and Factory Automation (ETFA). p. 1–8. IEEE, Limassol (Sep 2017). https://doi.org/10.1109/ETFA.2017.8247574
45. Van Rossum, G., Drake, F.L.: Python 3 Reference Manual. CreateSpace, Scotts Valley, CA (2009)
46. Wagner, S., Abdulkhaleq, A., Bogicevic, I., Ostberg, J.P., Ramadani, J.: How are functionally similar code clones syntactically different? An empirical study and a benchmark. PeerJ Computer Science **2**, e49 (Mar 2016). https://doi.org/10.7717/peerj-cs.49
47. Wang, P., Svajlenko, J., Wu, Y., Xu, Y., Roy, C.K.: Ccaligner: a token based large-gap clone detector. In: Proceedings of the 40th International Conference on Software Engineering. p. 1066–1077 (2018)
48. Wang, W., Li, G., Ma, B., Xia, X., Jin, Z.: Detecting code clones with graph neural network and flow-augmented abstract syntax tree. In: 2020 IEEE 27th International Conference on Software Analysis, Evolution and Reengineering (SANER). IEEE (feb 2020). https://doi.org/10.1109/saner48275.2020.9054857
49. Wei, H.H., Li, M.: Supervised deep features for software functional clone detection by exploiting lexical and syntactical information in source code. In: Proceedings of the 26th International Joint Conference on Artificial Intelligence. p. 3034–3040. IJCAI'17, AAAI Press, Melbourne, Australia (Aug 2017)
50. Zhang, J., Wang, X., Zhang, H., Sun, H., Wang, K., Liu, X.: A novel neural source code representation based on abstract syntax tree (may 2019). https://doi.org/10.1109/ICSE.2019.00086
51. Zhao, G., Huang, J.: DeepSim: Deep learning code functional similarity. In: Proceedings of the 2018 26th ACM Joint Meeting on European Software Engineering Conference and Symposium on the Foundations of Software Engineering. p. 141–151. ESEC/FSE 2018, Association for Computing Machinery, Lake Buena Vista, FL, USA (Oct 2018). https://doi.org/10.1145/3236024.3236068